\documentclass[a4paper]{article}

\usepackage{siunitx} 
\usepackage{graphicx} 
\usepackage{amsmath} 
\usepackage[hmargin={1.7cm,1.7cm},vmargin={1.3cm,2.4cm}]{geometry} 
\usepackage{cleveref}
\usepackage{subfigure}
\title{Smart Routing for Improved PLC Backhauling of the Radio Access Network} 

\author{Francesco Marcuzzi and Andrea M. Tonello\\
	Alpen-Adria-Universit\"at Klagenfurt\\
	Institute of Networked and Embedded Systems\\
	Universit\"atstrasse 65\\
	9020 Klagenfurt, Austria\\
	email: francesco.marcuzzi@aau.at; andrea.tonello@aau.at}

\date{} 

\begin{document}


\maketitle 

\begin{figure}[ht]
	\centering
	\includegraphics[scale = 0.3]{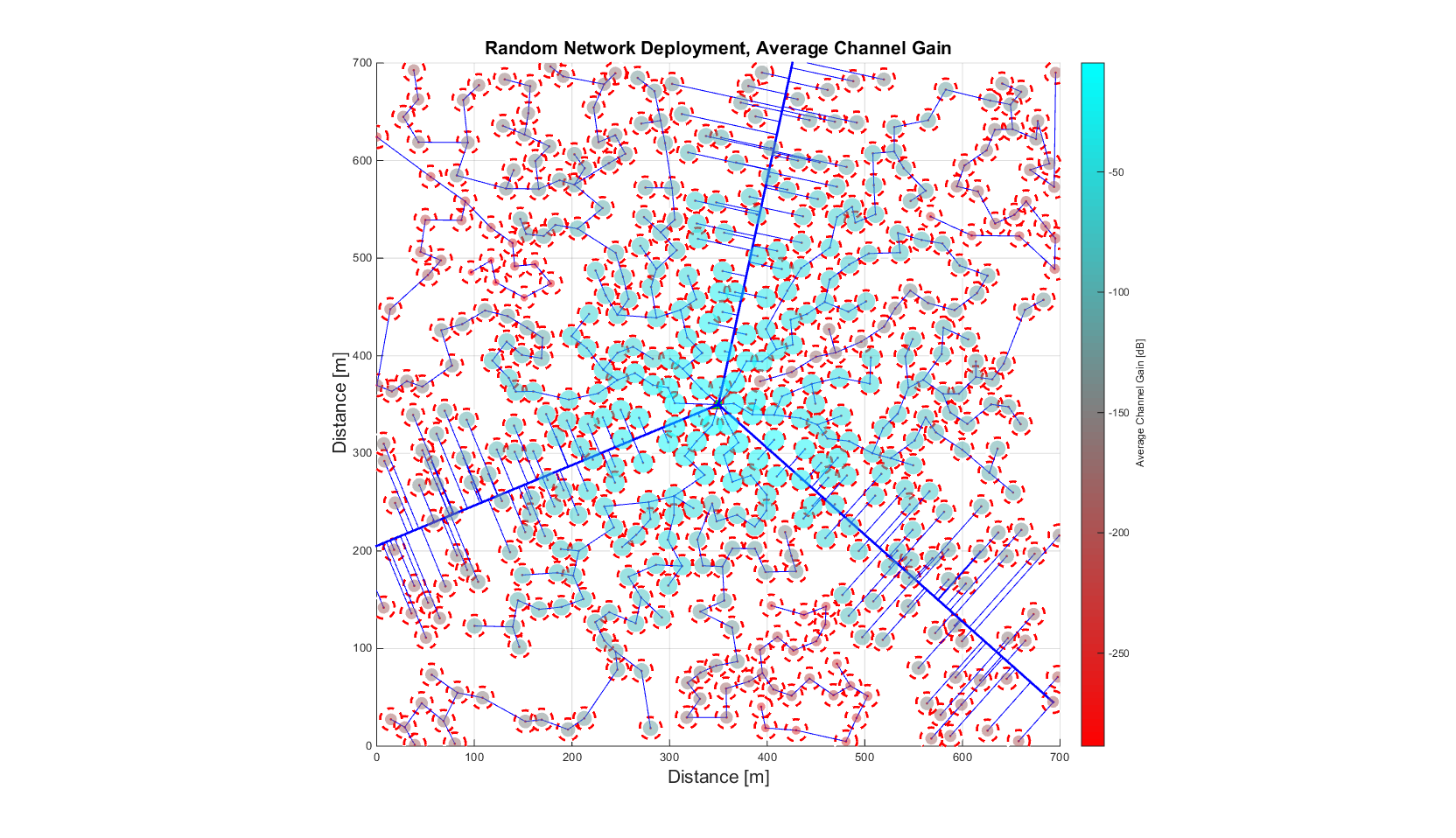}
	\caption{Color graded shades the average channel gain for each cell in a random deployed network.}
	\label{cfr}
\end{figure}

Nowadays, the population of connected cellular devices around the world is growing exponentially. In order to provide high-quality and seamless coverage, small-cell networks are being employed by telecommunication operators to increase spatial fragmentation and the reuse of cellular frequencies \cite{SCN}. The downside of this practice lies in the necessity for an additional step in the network's hierarchy, namely the back-haul section needed between the small cells and the Radio Access Network's (RAN) hub, located in the macro-cell's antenna. \\
Supporting the customers traffic requirements, reliability, easy deployment and maintenance and not exploiting end-user spectrum to avoid interference are basic ground rules to keep in mind when designing this back-haul section. Naturally, specific needs pertain to different application scenarios, which is the reason why there is no unique method to design an infrastructure apt to transfer data from small-cells to the RAN's hub. Both wired and wireless solutions are currently employed for this application, although neither of them offer the universally optimal solution, as there is a trade off between installation/maintenance costs, data rate capabilities and environmental factors to consider.\\
The power-line infrastructure offers an interesting asset for this application because of its wide-spread reach of devices. Power-line communications have not been widely considered for back-hauling purposes yet, although recent developments in modulation methods, media access control mechanisms, cooperative schemes and noise mitigation techniques have helped push the PLC performance \cite{PLC}. State-of-the-art broad band PLC technology offers very high speed data rates, in the order of 1 Gbps. Recent work by our group \cite{art, art2} has analyzed the possibility of employing PLC for this application, especially for scenarios where small-cells pertain to the femto-cell category, whose characteristics include very low power, short range and low number of supported users. This kind of cells allows for the most dense fragmentation and can be used for in-building applications as IoT networks. The results stemmed from this analysis revealed that requirements on PLC technology are in the ten of Mbps, when using a traffic generation model derived from realistic data and cellular network paradigms.\\
Through the inclusion of a transmission line theory model, it was possible to improve the previous results by considering a realistic channel response based on data retrieved from real cables. The model was based on the work presented in \cite{TV01,TV02}. The noise behavior, in this phase of the analysis, is assumed Gaussian, which is not realistic, but at least sets a bound in relation to a very well-known metric for communication channels. Each femto-cell is assumed as uniquely belonging to a private house, thus modeled as a resistance for the channel response computation. Once this is carried out, each channel response is employed to further compute the capacity of each link from the cell back to the CCo; \Cref{cfr} shows a random network deployment where the shades of color indicate the average quality of the channel through the gain metric. The average channel gain (ACG) is closely correlated to the effective capacity of the link, as in fact results for each density step show that a throughput in the order of 10 Mbps is achieved for ACGs of about -100 dB, while around -120 dBm the capacity goes below the 1 Mbps threshold.

\begin{figure}[ht]
	\centering
	\includegraphics[scale = 0.5]{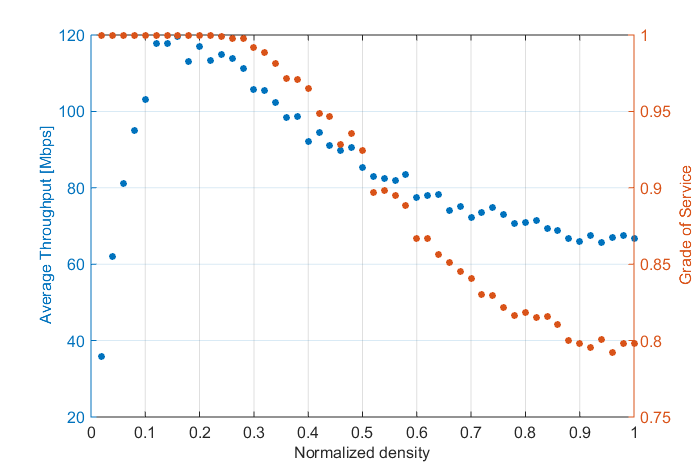}
	\caption{Evaluation metrics for the simulated network: effective throughput and grade of service.}
	\label{gm}
\end{figure}

Using the effective capacity computed through the TL theory model in combination with the aforementioned traffic generation model, it is possible to get deeper insight on the performance of the PLC technology. Assuming that the CCo operates seamlessly a resource allocation protocol, independently for each sector of the network, and that the media access scheme is a simple TDMA based on how much throughput each cell requires, we can establish a pair of metrics to show the trend of the network's behavior. First of all, we can show the average throughput achieved by a generic cell in the network based on the effective amount of data transmitted by each cell to the CCo. The second metric is the Grade of Service (GoS) which shows the average ratio between the effective throughput of the cells and the one they require to be completely satisfied by the connection. By running a large number of simulations, \Cref{gm} was obtained and it was possible to infer that effective throughput reaches its maximum value (around 120 Mbps) for a normalized density between 0.1 and 0.2 (when the network of femto-cells cover between 10 and 20 percent of the simulated geographical territory). Values of throughput outside this range tend to be lower, because of the scarcity of cells and traffic for lower densities and because of the lack of resource for higher ones. On the other hand, GoS shows a very steady, almost unitary trend for densities below 30 percent, after which starts linearly decreasing until it settles around 80 percent for completely covered territories.\\

\end{document}